\documentclass[pra,twocolumn,showpacs]{revtex4}
\usepackage{amssymb}
\usepackage{amsmath}
\usepackage{graphicx}

\newcommand{\bi}[1]{\textbf{\textit{#1}}}

\begin{document}

\title{Universality in tightly bound 3-boson systems}

\author{Y. Z. He$^2$}

\author{Y. Z. Fang$^2$}

\author{C. G. Bao$^{1,2,}$}

\thanks{Corresponding author: stsbcg@mail.sysu.edu.cn}
\affiliation{$^1$Center of Theoretical Nuclear Physics,
National Laboratory of Heavy Ion Accelerator, Lanzhou, 730000,
People's Republic of China}

\affiliation{$^2$State Key Laboratory of Optoelectronic
Materials and Technologies, School of Physics and Engineering,
Sun Yat-Sen University, Guangzhou, 510275, People's Republic of
China}

\begin{abstract}
The effects of two distinct operations of the elements of the
symmetry groups of a Hamiltonian on a quantum state might be
equivalent in some specific zones of coordinate space. Making
use of the matrix representations of the groups, the
equivalence leads to a set of homogeneous linear equations
imposing on the wave functions. When the matrix of the
equations is non-degenerate, the wave functions will appear as
nodal surfaces in these zones. Therefore, the equivalence leads
to the existence of inherent nodal structure in the quantum
states. In this paper, trapped 3-boson systems with different
types of interactions are studied. The structures of the
tightly bound eigenstates have been analyzed systematically.
The emphasis is placed to demonstrate the universality arising
from the common inherent nodal structures.
\end{abstract}


\maketitle


\subsection{Introduction}

It is well known that universality exists in weakly bound
3-body system, where the Efimov states
emerge.\cite{r_EV1,r_EV2} These states arise from the inherent
mysteries of quantum mechanics, and they are governed by a
universal law, not depend on the details of dynamics. In this
paper, another kind of universality that exists in tightly
bound quantum mechanic few-body systems is revealed and
confirmed numerically.

Usually, the Hamiltonian of identical particles is invariant
under the operations of a set of symmetry groups $G_{\alpha }$,
$G_{\beta }$, $\cdot\cdot$ (including the permutation
group).\cite{r_RDE,r_RG,r_LEM,r_EJP} Consequently, the
eigenstates $\{\Psi_i(X)\}$ are classified according to the
representations of these groups, where $X$ denotes a set of
coordinates and $i$ is a serial number. Let $g_{\alpha }$ be an
element of $G_{\alpha}$ , $g_{\beta}$ be that of $G_{\beta}$
(each of them may be an element of the direct product of
groups), and $\Xi$ denotes a special zone in the
high-dimensional coordinate space. When $X\in\Xi$, the
operations of $g_{\alpha}$ and $g_{\beta}$ might be equivalent
so that $g_{\alpha}\Psi_i(X)=g_{\beta}\Psi_i(X)$. For an
example, when $\Xi$ is the zone of regular triangles (RT),
$g_{\alpha}$ is the rotation about the normal of the triangle
by $2\pi/3$, and $g_{\beta}$ is a cyclic permutation of
particles, then $g_{\alpha}$ and $g_{\beta}$ are equivalent in
$\Xi $. Making use of the representations of groups, the
equivalence leads to the establishment of a set of homogeneous
linear equations $\sum_{i'}[D_{i'i}^{\alpha}(g_{\alpha})
-D_{i'i}^{\beta}(g_{\beta})]\Psi_{i'}(X)=0$ in $\Xi$, where
$D_{i'i}^{\alpha}(g_{\alpha})$ is the matrix element of the
representation. When the matrix of this set of equations is
non-degenerate, the set $\Psi_i(X)$ must be zero in $\Xi $. In
this case, $\Xi$ becomes a prohibited zone and the wave
function appears as an inherent nodal surface
(INS).\cite{r_BCG1} Since the matrixes of representations are
absolutely irrelevant to dynamics, the appearance of the INS is
universal disregarding the kind of systems (nuclear, atomic, or
molecular) and the details of dynamic parameters.

The inherent nodal structures of 3-boson systems (without a
trap) has been studied by the authors
previously.\cite{r_BCG2,r_BCG3} Accordingly, the quantum states
can be naturally classified according to their inherent nodal
structures.\cite{r_BCG2,r_BCG3,r_MT,r_PMD} Due to the progress
of techniques, a few atoms can be tightly bound in a magnetic
or optical trap. Thus the study of these systems is meaningful.
The present paper extends the previous research of the authors
and is dedicated to trapped 3-body systems. A more effective
and simpler approach is proposed for the analysis. In
particular, in addition to the RT, the symmetry constraint
associated with the isoceles triangle (IST) and the collinear
geometry (COL) has also been taken into account. By studying
the spectra, the root mean square radii, the particle spatial
distribution, the weights of $Q$-components ($Q$ is the
projection of the total orbital angular momentum $L$ on the
third body-axis), and the shape-densities, systematic knowledge
on the quantum states has been obtained. The emphasis is placed
on demonstrating the universality by comparing different
3-boson systems. In addition to boson systems, a short
discussion on fermion systems is also given at the end.

\subsection{Hamiltonian, spectra, and the eigen-states}

It is assumed that the three particles are confined by an
isotropic harmonic trap $\frac{1}{2}m\omega^2 r_i^2$. Due to
the trap numerous bound states exist, therefore a systematic
analysis can be made. $\hbar\omega$ and $\sqrt{\hbar/m\omega}$
are used as units of energy and length. Three types of
interactions are assumed:
$V_{\mathrm{A}}(r)=10(2e^{-(r/1.428)^2}-e^{-(r/2.105)^2})$,
$V_{\mathrm{B}}(r)=1000e^{-3r}-40/r^6$ ($r\geq 1.2$) and
$V_{\mathrm{B}}(r)=V_{\mathrm{B}}(1.2)$ ($r<1.2$), and
$V_{\mathrm{C}}(r)=0$ ($r\geq 1$) and $V_{\mathrm{C}}(r)=15$
($r<1$). The type $V_{\mathrm{A}}$ has a short-ranged character
and was previously used in nuclear physics for the
$\alpha$-particles, $V_{\mathrm{B}}$ belongs to the Van der
Waals type for atoms, while $V_{\mathrm{C}}$ is just a hard
core potential. In fact, the interactions are chosen quite
arbitrary, just to show the inherent universality among
different systems. The Hamiltonian is
\begin{eqnarray}
 H
 &=& \sum_i
     (-\frac{1}{2}
       \nabla_i^2
      +\frac{1}{2}
       r_i^2)
      +\sum_{i<j}
       V_J
       (|\bi{r}_{i}-\bi{r}_{j}|),  \nonumber \\
 (J &=&A,B,\text{ or }C).
 \label{e01_H}
\end{eqnarray}
A set of basis functions is introduced to diagonalize the
Hamiltonian to obtain the spectra and the eigenstates. The
details are given in the appendix. Note that both
$V_{\mathrm{A}}$ and $V_{\mathrm{B}}$ contain a minimum, and
therefore the low-lying states will pursue a better geometry so
that the particles are appropriately separated and the total
potential energy can be thereby lower. This pursuit is not
obvious for the case with $V_C$, where no minimum is contained.

\begin{figure}[tbp]
\centering \resizebox{0.9\columnwidth}{!}{
 \includegraphics{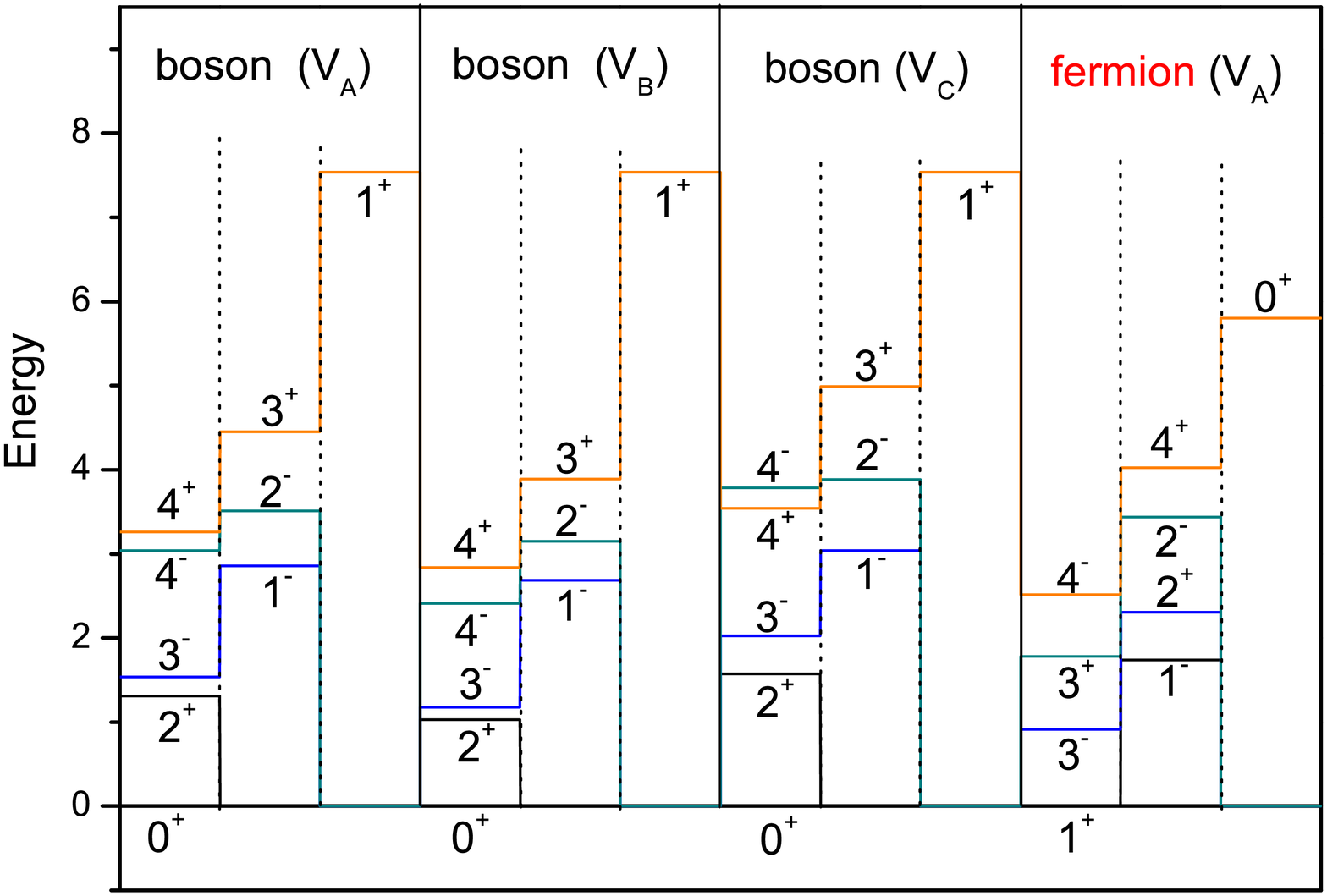} }
 \caption{Spectra of 3-body systems with $L=0$ to $4$ and parity
$\Pi=\pm 1$. Only the first-states are involved. For bosons
three types of interactions $V_{\mathrm{A}}$, $V_{\mathrm{B}}$,
and $V_{\mathrm{C}}$ are considered. For fermions (the spatial
wave functions are totally antisymmetric) only $V_{\mathrm{A}}$
is considered. The levels have been shifted so that all the
ground states have energy zero. The energy unit for the two
cases with $V_{\mathrm{A}}$ is $\hbar\omega$. In the case with
$V_{\mathrm{B}}$ or $V_{\mathrm{C}}$ the energy unit has been
redefined so that the energy of the $1^+$ state has the same
value as the one with $V_{\mathrm{A}}$. In every case the
states are divided into three groups. For bosons, when
different interactions are used the members and sequence in
each group are the same (the only exception is that $4^-$ is
higher than $4^+$ when $V_{\mathrm{C}}$ is used). }
 \label{fig1}
\end{figure}

When the c.m. motion is removed an eigenstate is denoted as
$\Psi_{L,M}^{\Pi ,i}$, where $M$ is the $Z$-component of $L$,
$\Pi$ the parity, and $i$ denotes the $i$-th state of a
$L^{\Pi}$-series. The $i=1$ state (the lowest in a series) is
called a first-state. The label $i$ is usually ignored when
$i=1$. For the three types of interaction the spectra of nine
first-states are plotted in Fig.~\ref{fig1} (the $L^{\Pi}=0^-$
state does not exist due to the "Rule 1" given below), where
the spectrum at the right is for fermions and will be concerned
later. It turns out that the three spectra at the left for
bosons are similar. The levels can be similarly divided into
three groups. In a group, the structures of the states are
similar as shown below, and those with a larger $L$ are higher
due to the rotation energy. Many common features existing in
the three spectra (say, $1^+$ is particularly high, $3^+$ is
much higher than $3^-$, and $2^-$ is much higher than $2^+$,
etc.) can not be explained simply by dynamics. It implies the
existence of a more fundamental reason.

\begin{figure}[tbp]
\centering \resizebox{0.9\columnwidth}{!}{
 \includegraphics{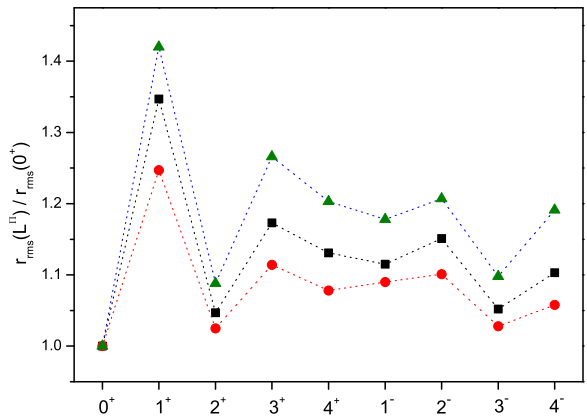} }
 \caption{The ratio
$r_{\mathrm{rms}}(L^{\Pi})/r_{\mathrm{rms}}(0^+)$, of the
$L^{\Pi}_1$ states. Square is for $V_{\mathrm{A}}$, circle for
$V_{\mathrm{B}}$, and triangle for $V_{\mathrm{C}}$. }
 \label{fig2}
\end{figure}

The root mean square radius $r_{\mathrm{rms}}(L^{\Pi})$ is
extracted from $\Psi_{L,M}^{\Pi}$ to evaluate the size. The
ratios $r_{\mathrm{rms}}(L^{\Pi})/r_{\mathrm{rms}}(0^+)$ are
shown in Fig.~\ref{fig2} where the three dotted lines guiding
the eyes go up and down in a synchronous way. For all the three
types of interactions, the sizes of $0^+$, $2^+$, and $3^-$ are
relatively smaller, while the sizes of $1^+$ is the largest.
From dynamics, the $4^+$ and $4^-$ states are expected to be
larger in size because the particles are pushing out by a
stronger centrifugal force, but in fact not. It demonstrates
again the existence of a more fundamental reason.

Let $\bi{k}'$ be a unit vector vertical to the plane of
particles. For $L\neq 0$ states, the relative orientation of
$\bi{k}'$ and $L$ is a noticeable feature. Let a body frame
$\Sigma'$ with the base-vectors $\bi{i}'-\bi{j}'-\bi{k}'$ be
introduced, where $\bi{j}'$ lies along $\bi{R}
\equiv\bi{r}_3-(\bi{r}_1+\bi{r}_2)/2$. Then, the eigenstates
can be expressed as $\Psi _{LM}^{\Pi}(X)=\sum_Q D_{QM}^L(\Omega
)\Psi _{LQ}^{\Pi }(X')$ where $X$ denotes the set of
coordinates relative to a fixed frame $\Sigma$, and $X'$
denotes the set relative to $\Sigma'$, $\Omega$ denotes the
three Euler angles responsible for the transformation from
$\Sigma'$ to $\Sigma $.\cite{r_EAR} Starting from the normality
$1=\langle\Psi_{L,M}^{\Pi}|\Psi_{L,M}^{\Pi}\rangle$, we carry
out first the integration over $\Omega$. Then we have
\begin{eqnarray}
 1
 &=& \sum_Q
     \frac{8\pi^2}{2L+1}
     \int \mathrm{d}X' |
     \Psi_{L,Q}^{\Pi}(X') |^2
 \equiv
     \sum_Q
     W_Q,
 \label{e02_1}
\end{eqnarray}
where $\mathrm{d}X'^2
R^2\mathrm{d}r\mathrm{d}R\sin\theta\mathrm{d}\theta$, $\theta$
is the angle between $\bi{r}\equiv \bi{r}_2-\bi{r}_1$ and
$\bi{R}$. $W_Q$ is the weight of the $Q$-component
$\Psi_{LQ}^{\Pi}$. Note that $Q$ is the projection of $L$ along
$\bi{k}'$. Thus, in $\Psi_{LL}^{\Pi}$, $L$ is essentially lying
along $\bi{k}'$ and therefore is essentially vertical to the
plane of particles. Whereas in $\Psi_{L0}^{\Pi}$, $L$ is
essentially lying on the $\bi{i}'-\bi{j}'$ plane and therefore
is coplanar with the particles. Thus, the relative orientation
between $L$ and $\bi{k}'$ can be understood from $W_Q$.

Carrying out the integration over $\mathrm{d}X'$ numerically,
$W_Q$ can be obtained. Since $W_Q=W_{-Q}$, we define
$\bar{W}_Q=2W_Q$ (if $Q\neq 0$), or $=W_Q$ (if $Q=0$). They are
given in Tab.~\ref{tab1}.

\begin{table}[htbp]
\caption{The weights of the $|Q|$-components obtained by using $V_A$.}%
\begin{ruledtabular}
  \label{tab1}
  \begin{tabular}{llllll}
                &   $\bar{W}_0$     &   $\bar{W}_1$     &   $\bar{W}_2$     &   $\bar{W}_3$     &   $\bar{W}_4$     \\
  \hline
  $0^+$         &   1               &                   &                   &                   &                   \\
  $1^+$         &   1               &                   &                   &                   &                   \\
  $2^+$         &   0.862           &                   &   0.138           &                   &                   \\
  $3^+$         &   0.001           &                   &   0.999           &                   &                   \\
  $4^+$         &   0.529           &                   &   0.305           &                   & 0.166             \\
  $1^-$         &                   &   1               &                   &                   &                   \\
  $2^-$         &                   &   1               &                   &                   &                   \\
  $3^-$         &                   &   0.016           &                   &   0.984           &                   \\
  $4^-$         &                   &   0.016           &                   &   0.929           &
 \end{tabular}
 \end{ruledtabular}
\end{table}

$\bar{W}_Q$ are far from uniform but concentrated in one
$|Q|-$component (except in $4^+$). When other interactions are
used the qualitative features of $\bar{W}_Q$ remain unchanged
(say, for $V_A$, $V_B$ and $V_C$, respectively, $3^-$ has
$\bar{W}_3=0.984$, 0.991, and 0.972). Thus, again, there is a
fundamental reason beyond dynamics.

\subsection{Symmetry constraint and the classification of states}

To clarify the fundamental reason, we study the effect of
equivalent operations. Let the operator of a rotation about an
axis $\bi{a}$ by an angle $\beta$ be denoted as
$\mathcal{R}_{\beta}^{\bi{a}}$. When $\bi{a}$ is $\bi{k}'$
which is vertical to the plane of particles, the rotation by
$\pi$ is equivalent to an inversion $\mathcal{I}$ with respect
to the c.m.. Thus we have
$\mathcal{R}_{\pi}^{\bi{k}'}\Psi_{LQ}^{\Pi}=e^{-i\pi Q}\
\Psi_{LQ}^{\Pi}=\Pi\Psi_{LQ}^{\Pi}$. It leads to the "Rule 1:
\textit{only $Q$ even (odd) components are allowed for parity
even (odd) states}" as shown in the table. Furthermore, at a
RT, $\mathcal{R}_{2\pi/3}^{\bi{k}'}\Psi_{LQ}^{\Pi}
=e^{-i\frac{2\pi}{3}Q}\
\Psi_{LQ}^{\Pi}=P_{\mathrm{cyc}}\Psi_{LQ}^{\Pi}$, where
$P_{\mathrm{cyc}}$ denotes a cyclic permutation of particles
and will cause no effect. Thus, we have the "Rule 2:
\textit{$\Psi_{L,Q}^{\Pi}$ is nonzero at a RT only if $Q=0,\pm
3,\pm 6,\cdots$".} When the three particles turn out to form an
IST (including the RT as a special case) with the symmetric
axis $\bi{v}$ lying on the $\bi{i}'$-$\bi{j}'$ plane, we have
the equivalence $\mathcal{R}_{\pi
}^{\bi{v}}\Psi_{LQ}^{\Pi}=P_{ij}\Psi_{LQ}^{\Pi}$, where
$P_{ij}$ denotes the interchange of the two particles at the
bottom of the IST. It leads to $e^{-i2\delta Q}(-1)^{L+Q}\
\Psi_{L,-Q}^{\Pi}=\varepsilon\Psi_{LQ}^{\Pi}$, where $\delta$
is the angle between $\bi{v}$ and $\bi{j}'$, and
$\varepsilon=\pm 1$ depends on the statistics. Thus, for bosons
and for $Q=0$, we have the "Rule 3: \textit{$\Psi_{L,0}^{\Pi}$
is nonzero at an IST only if $L$ is even}". With the three
rules, when $L\leq 4$, the RT- accessible components are
$\Psi_{L=\mathrm{even},0}^{\Pi =1}$ and
$\Psi_{L=\mathrm{odd},\pm 3}^{\Pi=-1}$.

Since all the three bonds can be optimized in a RT, the
RT-accessible (RT-ac) components are much favored by the lower
states. Among the nine states under consideration, the RT-ac
components can be contained in the $0^+$, $2^+$, $4^+$, $3^-$
and $4^-$ states. They are called the RT-ac states and are
dominated by the RT-ac component with $|Q|=0$ (3) if $\Pi=1$
($-1$) (as shown in Table 1), thereby their energies can be
much lower and they constitute the lowest group in the
spectrum. Their wave functions are distributed surrounding a RT
resulting in having a smaller size. The domination of the RT-ac
component explains why $\bar{W}_0$ or $\bar{W}_3$ is
particularly large in the RT-ac states.

For $1^+$, the $|Q|=1$ component is prohibited by "Rule 1",
while in the $Q=0$ component the IST is prohibited by "Rule 3".
The prohibition of all the IST causes serious consequence. It
implies that the wave function is expelled from the important
zones of coordinate space (see below) resulting in having a
very high energy and a large size. This explains why this state
belongs to the highest group in the spectra.

For $1^-$ and $2^-$, only $|Q|=1$ is allowed due to "Rule 1".
This component is IST-ac but RT-inaccessible (RT-inac).
Accordingly, these states are classified to the second group.
For $3^+$, both $|Q|=0$ and $2$ components are allowed. Due to
"Rule 3", the former is IST-inac and therefore extremely
unfavorable. Thus the weights are greatly concentrated in
$\bar{W}_{2}$ as shown in Table 1. Since the $|Q|=2$ component
is also IST-ac and RT-inac, $3^{+}$ belongs also to the second
group.

Since \textit{the classification} is based on the accessibility of the RT
and IST originating from symmetry constraint, it \textit{is universal for
all kinds of 3-boson systems}.

\subsection{One-body densities}

\begin{figure}[tbp]
\centering \resizebox{0.9\columnwidth}{!}{
 \includegraphics{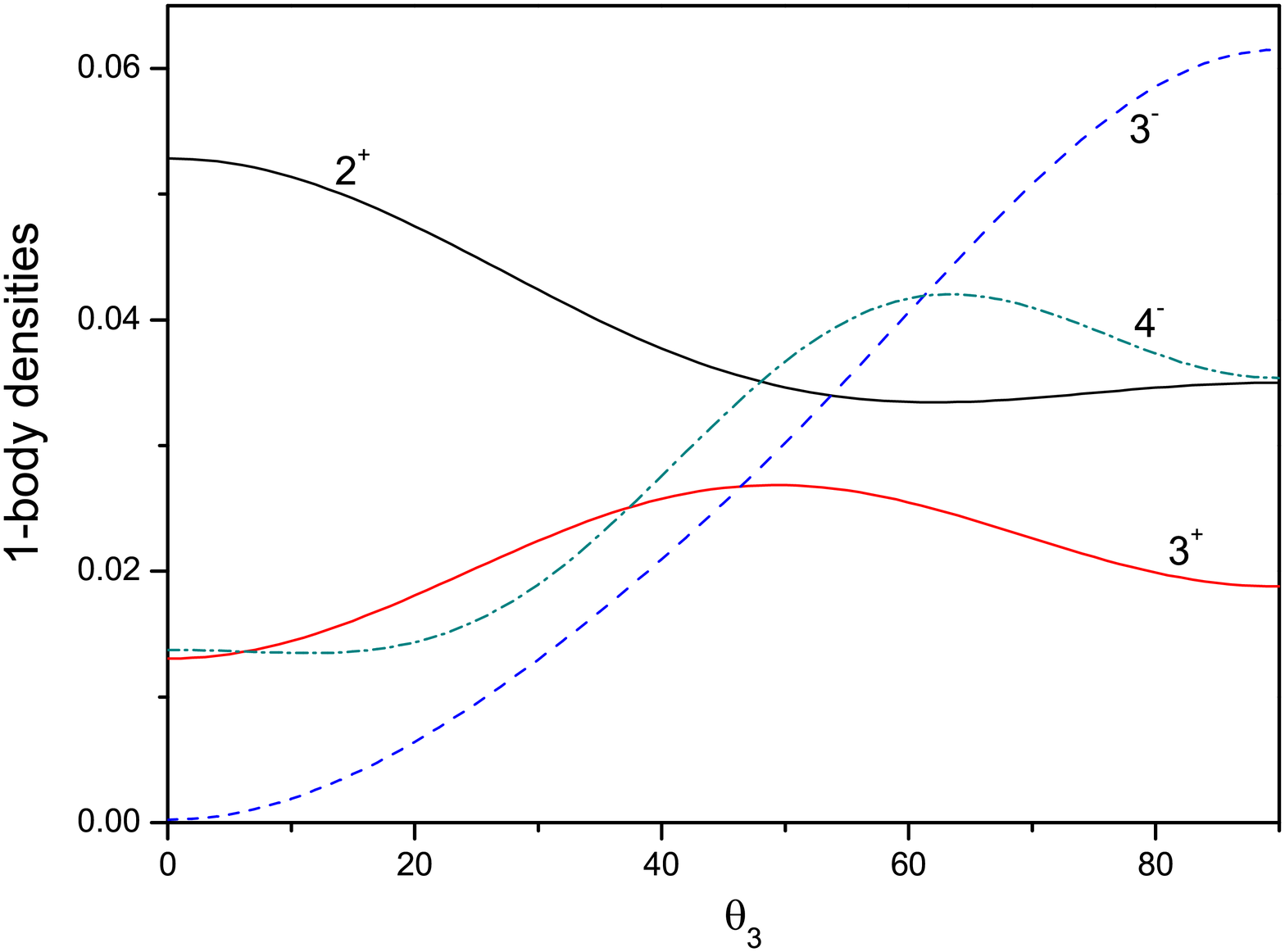} }
 \caption{One-body densities $\rho_1(r_3,\theta_3)$ of the
$\Psi_{LL}^{\Pi}$ states calculated by using $V_A$ and with
$r_3$ given at $r_{\mathrm{rms}}$. $\theta_3$ is the polar
angle. $\rho_1(r_3,\theta_3)=\rho_1(r_3,\pi-\theta_3)$, and
$\rho_1$ does not depend on the azimuthal angle.}
 \label{fig3}
\end{figure}

Since the weights $\bar{W}_Q$ are related to the relative
orientation between $\bi{k}'$ and $L$, these weights are
related to the one-body densities $\rho_1$ which is an
observable. Let us go back to the fixed-frame and deal with the
eigenstates $\Psi_{LM}^{\Pi}(\bi{r},\bi{R})$. From the
normality, we have
\begin{eqnarray}
 1
  =  \int
     \mathrm{d}\bi{r}
     \mathrm{d}\bi{R} |
     \Psi_{LM}^{\Pi} |^2
  \equiv
     \int
     \mathrm{d}\bi{r}_3
     \rho_1(\bi{r}_3),
 \label{e03_1}
\end{eqnarray}
where $\bi{r}_3=\frac{2}{3}\bi{R}$ is the position vector of
the third particle relative to the c.m., and
$3\rho_1(\bi{r}_3)$ is the probability density of finding a
particle (disregarding which one) at $\bi{r}_3$. Let $M=L$, it
implies that $L$ is given along the $Z$-axis. The $\rho_1$
extracted from $\Psi_{LL}^{\Pi }$ is plotted in
Fig.~\ref{fig3}. For $3^-$, since this state is dominated by
$\bar{W}_3$, $\bi{k}'$ is essentially lying along $L$ (i.e.,
along the $Z$-axis at the choice $M=L$). Therefore, it is
expected that the $\rho_1$ of $3^-$ would be distributed close
to the X-Y plane. This presumption is confirmed in
Fig.~\ref{fig3}. For $4^-$, $\bar{W}_3$ is also dominant, thus
the direction of $L$ is close to but deviates a little from the
Z-axis. Accordingly, the peak of $\rho_1$ of $4^-$ is close to
but deviates from $\theta_3=90^{\circ}$. $3^+$ is dominated by
$\bar{W}_2$ . Accordingly, the peak is farther apart from
$90^{\circ}$. Whereas $2^+$ is dominated by $\bar{W}_0$.
Therefore $\bi{k}'$ is essentially vertical to the $Z$-axis,
thus the plane of the particles prefer to be coplanar with the
Z-axis if $|M|=L$, and the distribution is no more
close to the equator. In fact, the $\rho_1$ of $2^+$ is peaked at $%
\theta_3=0 $.

Although Fig.~\ref{fig3} is from $V_A$, the feature is common
for various 3-boson systems. For examples, when $V_A$, $V_B$,
and $V_C$ are used, the $\rho_1$ of $4^-$ is peaked at
$63^{\circ}$, $63^{\circ}$, and $66^{\circ}$, respectively, and
the $\rho_1$ of $3^+$ is peaked at $50^{\circ}$, $49^{\circ}$,
and $49^{\circ}$.

\subsection{Shape-densities}

Let us study the structures of the states in more detail in the
body-frame. We define the hyper-radius
$\mathfrak{h}=\sqrt{\frac{1}{2}r^2+\frac{2}{3}R^2}$. This
quantity is invariant under the transformation between
different sets of Jacobi coordinates. Therefore, it is suitable
to describe the size. Furthermore, we define $\tan\beta
=\sqrt{2/3}R/(\sqrt{1/2}r)$. When $r$ and $R$ are replaced by
$\mathfrak{h}$ and $\beta $ as arguments, Eq.~(\ref{e02_1}) can
be rewritten as
\begin{equation}
 1
  =  \int
     \mathrm{d}\mathfrak{h}
     \mathrm{d}S\
     \rho_{\mathrm{sha}}(X'),
\end{equation}
where $\mathrm{d}S
=\frac{3\sqrt{3}}{(1+\tan^2\beta)(1+1/\tan^2\beta)}
\mathrm{d}\beta\sin\theta\mathrm{d}\theta$ is an infinitesimal
deformation, and
\begin{equation}
 \rho_{\mathrm{sha}}(X')
  =  \frac{8\pi^2}{2L+1}
     \mathfrak{h}^5
     \sum_Q |
     \Psi_{LQ}^{\Pi}(X')|^2,
\end{equation}
is the probability density that the system has a given size and
a given shape (while an integration over the orientation of the
shape has already been carried out). Therefore
$\rho_{\mathrm{sha}}(X')$ is called the shape-density which
relates directly to the structure.

\begin{figure}[tbp]
\centering \resizebox{0.9\columnwidth}{!}{
 \includegraphics{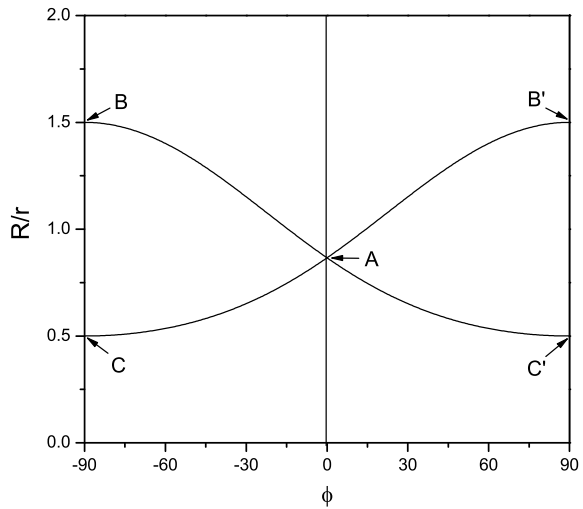} }
 \caption{Geometries in the $(\phi,R/r)$ subspace. Where both
the curves $\widetilde{BAC'}$ and $\widetilde{B'AC}$ correspond
to the IST. These curves fulfill the equation
$R/r=\frac{1}{2}\sqrt{5-4F}$, and
$F=\frac{1}{2}(\cos^2\phi\pm\sqrt{(3+\sin^2\phi)\sin^2\phi})$.
The vertical line $\phi=0$ corresponds also to the IST. The
intersection $A$ ($\phi=0$, $R/r=\sqrt{3}/2$) corresponds just
to a RT. Both the vertical lines $\phi=\pm 90^{\circ}$
correspond to collinear structures COL. In particular, the
point $B$ and $B'$ together with the horizontal line $R/r=0$
correspond to a symmetric COL, namely, a particle sits at the
middle of the other two. The point $C$ and $C'$ correspond also
to a COL but with two particles overlap with each other. }
 \label{fig4}
\end{figure}

Let $\phi$ be the azimuthal angle of $\bi{r}$ lying on the
$\bi{i}'$-$\bi{j}'$ plane. Since $\bi{R}$ is given lying along
$\bi{j}'$, $\phi=\pi/2-\theta$. Let $\mathfrak{h}$ be given at
$\sqrt{3}r_{\mathrm{rms}}$ (Let $r_i$ be the distance of the
$i$-th particle from the c.m.. Then
$\mathfrak{h}\equiv\sqrt{r_1^2+r_2^2+r_3^2}$ , and the average
$\bar{\mathfrak{h}}=\langle|r_1^2+r_2^2+r_3^2|\rangle^{1/2}
=\sqrt{3}r_{\mathrm{rms}}$). Examples of $\rho_{\mathrm{sha}}$
as a function of $\phi$ and
$R/r\equiv\frac{\sqrt{3}}{2}\tan\beta$ are given below. The
geometry associated with the arguments $(\phi,R/r)$ is shown in
Fig.~\ref{fig4}.

\begin{figure}[tbp]
\centering \resizebox{0.9\columnwidth}{!}{
 \includegraphics{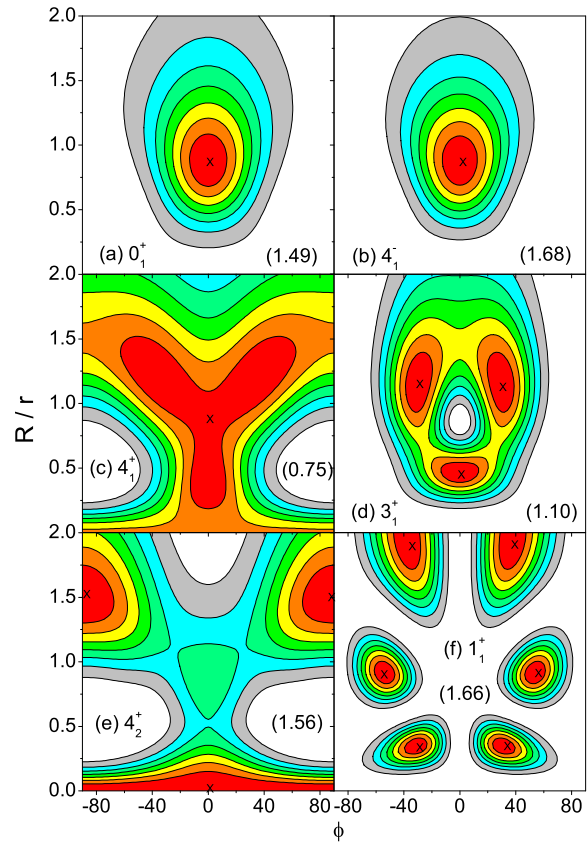} }
 \caption{The shape-densities $\rho_{\mathrm{sha}}$ of some
selected $L_1^{\Pi}$ first-states (an exception is \ref{fig5}e
for the second-state) obtained by using $V_A$. For each state
the hyper-radius $\mathfrak{h}$ is given at
$\sqrt{3}r_{\mathrm{rms}}$, and the orientation of the shape
has been integrated. The maxima of $\rho_{\mathrm{sha}}$ are
marked by a cross and their values are given inside the
parenthesis in each panel. The values of the contours form an
arithmetic series and decrease to zero. }
 \label{fig5}
\end{figure}

When $V_A$ is used, the $\rho_{\mathrm{sha}}$ of the RT-ac
first-states are very similar with each other (except $4^+$) as
shown in Fig.~\ref{fig5}a and \ref{fig5}b. The
$\rho_{\mathrm{sha}}$ of $4^+$ is shown in \ref{fig5}c. All of
them are peaked at $(\phi,R/r)=(0,\sqrt{3}/2)$ associated with
a RT as expected. In 5c the peak is extended along three
directions each leading to a COL. All the three RT-inac and
IST-ac first-states are similar as shown in \ref{fig5}d, where
a node appears at the RT and the three peaks are all associated
with a flattened IST with particles 3, 1, and 2 at the top,
respectively. The top angles of the IST are $94^{\circ}$,
$101^{\circ}$, and $105^{\circ}$, respectively for $3^+$,
$1^-$, and $2^-$. The $\rho_{\mathrm{sha}}$ of the IST-inac
state $1^+$ is shown in \ref{fig5}f, where a number of nodal
lines exist in the $(\phi ,R/r)$ subspace, and the wave
functions are expelled and compressed into six separate
regions. The appearance of so many nodal lines implies the
existence of a very strong inherent oscillation. This explains
why $1^+$ is very high in energy. From the different patterns
of the $\rho_{\mathrm{sha}}$, we know that the classification
based on the accessibility of geometries is valid.

For a COL, let $\bi{k}'$ be set along the line of particles,
then the rotation about $\bi{k}'$ by any angle causes nothing.
Hence the COL exists only in the $Q=0$ component (it implies
that $L$ is essentially vertical to $\bi{k}'$). Besides, a
rotation about $\bi{j}'$ by $\pi$ is equivalent to an
inversion. This leads to
$(-1)^L\Psi_{L0}^{\Pi}=\Pi\Psi_{L0}^{\Pi}$. Accordingly, only
the states with $(-1)^{L}=\Pi $ are allowed to have the COL.
Furthermore, for a symmetric COL, an inversion is equivalent to
an interchange of the two particles at the ends. Thus, the
symmetric COL exists only in positive parity states (for
bosons). When a COL is accessible but the symmetric COL is not,
the COL is not stable because an inherent node is contained and
a linear oscillation is inherently excited. On the other hand,
from pure dynamics, a COL will have a larger moment of inertia.
Comparing with the RT, a COL will have a higher interaction
energy but a smaller rotation energy. Hence, this structure
will be preferred by the states with a larger $L$ so that the
rotation energy can be considerably reduced. With this
analysis, among the nine states, obviously, the COL will be
more preferred by $4^+$. This explains the extension found in
\ref{fig5}c, but not in \ref{fig5}a and \ref{fig5}b. It is
expected that the COL will definitely appear in $i\geq 2$
states as shown in \ref{fig5}e for $4_2^+$, where the most
probable shape is a symmetric COL. Incidentally, $1^+$ as shown
in 5f is not only IST-inac but also COL-inac, this explains why
so many nodal lines emerge.

\begin{figure}[tbp]
\centering \resizebox{0.9\columnwidth}{!}{
 \includegraphics{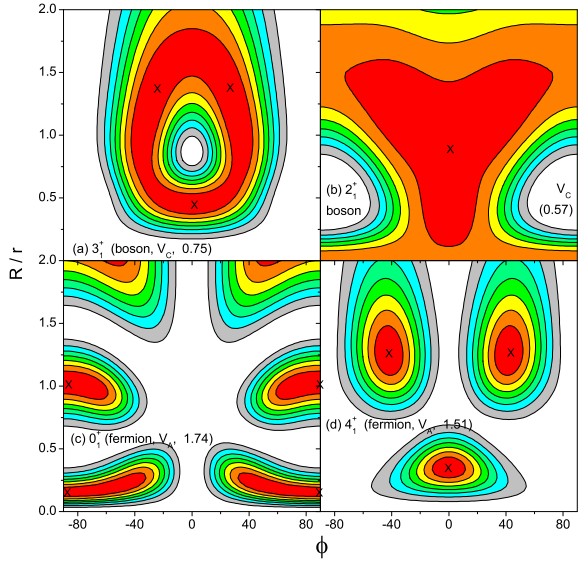} }
 \caption{ $\rho_{\mathrm{sha}}$ obtained by using $V_C$ for
bosons (a and b), and $\rho_{\mathrm{sha}}$ obtained by using
$V_A$ for fermions (c and d). Refer to Fig.~\ref{fig5}. }
 \label{fig6}
\end{figure}

When $V_B$ is used, the qualitative features of Fig.~\ref{fig5}
remain unchanged. Note that all the first states will do their
best to pursue a structure favorable in binding and free from
the symmetry constraint. This leads to the similarity. When
$V_C$ is used, the total potential energy is not sensitive to
the details of geometry because the interaction does not
contain a minimum. Hence, the pursuit to a better geometry is
less anxious. This is shown in \ref{fig6}a (Comparing with
\ref{fig5}d, the IST is no more well kept) and in \ref{fig6}b
(where the distribution extends towards the COL. This tendency
is not explicit in $2_1^+$ if $V_A$ or $V_B$ is used).

\subsection{Fermion systems}

For fermion systems, we consider the case that the spatial wave
function is totally antisymmetric against particle permutation.
Then the above Rule 1 and 2 hold exactly while Rule 3 becomes
\textit{"$\Psi_{L,0}^{\Pi}$ is nonzero at an $\mathrm{IST}$
only if $L$ is odd"}. With the three rules of fermions, when
$L\leq 4$, the RT- ac states are $1^+$, $3^+$, $3^-$, and
$4^-$. The RT-inac but IST-ac states are $1^-$, $2^+$, $2^-$,
and $4^+$. The IST-inac state is $0^+$. The spectrum from
numerical calculation by using $V_A$ is shown at the right of
Fig.~\ref{fig1}, where the classification into three groups is
revealed. As bosons, only the states with $(-1)^L=\Pi$ are
allowed to access the COL. However, the symmetric COL is
accessible only to $\Pi=-1$ states. From an analysis of the
$Q$-components, we know that $3^-$ has its RT lying on the
$i'$-$j'$ plane, while the RT of $3^+$ is upstanding.
Furthermore $3^-$ is symmetric-COL-ac, while $3^+$ is not. This
explains why $3^-$ is lower than $3^+$ even though they belong
to the same group. Similarly, the IST in $2^+$ is lying, that
in $2^-$ is not (due to the prohibition of $Q $ even
component). Besides, $2^+$ is COL-ac while $2^-$ is not. This
explains why $2^+$ is lower than $2^-$ even though they belong
to the same group.

Examples of $\rho_{\mathrm{sha}}$ of the 3-fermion states are
shown in Fig.~\ref{fig6}c and \ref{fig6}d. Fig.~\ref{fig6}c is
similar to \ref{fig5}f where the wave function is absolutely
expelled from the lines of IST. However, $0^+$ of fermions is
COL-ac (but symmetric COL-inac) while $1^+$ of bosons is not.
Therefore, under the same interaction the $0^+$ of fermions is
lower (see Fig.~\ref{fig1}), and its most probable shape is a
non-symmetric COL with an inherent strong linear oscillation
(due to the nodal lines). Comparing Fig.~\ref{fig6}d and
\ref{fig5}c (both for $4^+$ but for fermions and bosons,
respectively), the wave function of the former is expelled from
the RT and also from the symmetric COL, while the latter is
attracted by the COL. Thus the former is higher.

\subsection{Final remarks}

In conclusion, the features of the tightly bound 3-boson states
have been studied systematically. The universal existence of
inherent nodal structures has been demonstrated analytically
and numerically. The existence is irrelevant to the strength of
the trap. The quantum states belonging to the same set of
representations of the symmetry groups are constrained in the
same way disregarding the kind of particles. Since the
constraint arising from symmetry is absolutely not violable,
the effect of symmetry overtakes dynamics and is decisive. In
particular, there are prohibited zones (points, lines, ...) in
the coordinate space not accessible to specific states. This
leads to the establishment of a universal classification scheme
based on the accessibility of RT, IST, and COL, and leads to
the extensive similarity among different systems. Although this
paper is essentially dedicated to boson systems, universality
exists also in fermions systems due to the decisive role of
symmetry constraint.

\begin{acknowledgments}
The support from the NSFC under the grant No.10874249 is appreciated.
\end{acknowledgments}

\section*{Appendix}

\section{Diagonalization of the Hamiltonian and an evaluation of the accuracy%
}

When the c.m. coordinates have been removed, the Hamiltonian
for internal motion $H_{\mathrm{int}}$ can be expressed by a
set of Jacobi coordinates $\bi{r}=\bi{r}_2-\bi{r}_1$, and
$R=\bi{r}_3-(\bi{r}_1+\bi{r}_2)/2$. Let
\begin{equation}
 h(\mu,\bi{r})
  \equiv
    -\frac{1}{2\mu}
     \nabla_{\bi{r}}^2
    +\frac{1}{2}
     \mu r^2,
 \label{e_A1}
\end{equation}
which is the Hamiltonian of harmonic oscillation, Then
\begin{equation}
 H_{\mathrm{int}}
  =  h(1/2,\bi{r})
    +h(2/3,\bi{R})
    +\sum_{i<j}
     V(| \bi{r}_j
        -\bi{r}_i|).  \label{e_A2}
\end{equation}

Let us introduce a variational parameter $\gamma$. The
eigenstates of $h(\gamma,\bi{r})$ are denoted as
$\phi_{nl}^{\gamma }(\bi{r})$ with eigenenergy $2n+l+3/2$,
where $n$ and $l$ are the radial and angular quantum numbers,
respectively. From $\phi_{nl}^{\gamma }(\bi{r})$ a set of basis
functions for the 3-body system is defined as
\begin{equation}
 \Phi_{k,\Pi LM}^{\gamma }(123)
  \equiv
    ( \phi_{n_a l_a}^{\gamma/2}(\bi{r})
      \phi_{n_b l_b}^{2\gamma /3}(\bi{R}))_{LM},
 \label{e_A3}
\end{equation}%
where $l_a$ and $l_b$ are coupled to the total orbital angular
momentum $L$ and its $Z$-component $M$, $\Pi =(-1)^{l_a+l_b}$
the parity, $k$ denotes $n_a$, $l_a$, $n_b$, and $l_b$. These
functions should be further (anti)symmetrized, thus we define
\begin{equation}
 \tilde{\Phi}_{k,\Pi LM}^{\gamma}
  \equiv
     \sum_p
     \Phi_{k,\Pi LM}^{\gamma}(p_1 p_2 p_3),
 \label{e_A4}
\end{equation}
where the right side is a summation over the permutations.
Making use of the Talmi-Moshinsky coefficients,
\cite{r_BTA,r_BM,r_TW} each term at the right can be expanded
in terms of $\Phi_{k',\Pi LM}^{\gamma}(123)$. Say,
\begin{equation}
 \Phi_{k,\Pi LM}^{\gamma}(132)
  =  \sum_{k'}
     A_{kk'}^{\Pi L}
     \Phi_{k',\Pi LM}^{\gamma}(123),
 \label{e_A5}
\end{equation}
where the Talmi-Moshinsky coefficients $A_{kk'}^{\Pi L}$ can be
obtained by using the method given in the ref.\cite{r_TW}. Note
that the set $\{\tilde{\Phi}_{k,\Pi LM}^{\gamma}\}$ has not yet
been orthonormalized. Thus a standard procedure is further
needed to transform $\{\tilde{\Phi}_{k,\Pi LM}^{\gamma }\}$ to
a orthonormalized set $\{\tilde{\tilde{\Phi}}_{q,\Pi
LM}^{\gamma}\}$, where $q$ is a serial number. Each
$\tilde{\tilde{\Phi}}_{q,\Pi LM}^{\gamma}$ can be expanded in
terms of $\{\Phi_{k,\Pi LM}^{\gamma}(123)\}$. Finally, the set
$\{\tilde{\tilde{\Phi}}_{q,\Pi LM}^{\gamma}\}$ is used to
diagonalize $H_{\mathrm{int}}$. Due to the symmetrization the
matrix element of any pair of interaction is equal to that of
the pair 1 and 2. Thus only $\bi{r}$ and $\bi{R}$ are involved
in the calculation. This leads to a great convenience. When
$\Pi $ and $L$ are given, a series of states will be obtained
after the diagonalization. The lowest one of the series is
called the first state. Its energy should be minimized by
adjusting $\gamma$.

Due to having a trap, the bound states are highly localized,
the above basis functions are suitable and will lead to a good
convergency. Let $N_{ab}=2(n_a+n_b)+l_a+l_b$. The number of
basis functions is constrained by $N_{ab}\leq N_{max}$, the
latter is given. As an example for the bosons, the energy of
the lowest $0^+$ state with the interaction $V_A$ would be
2.06561, 2.06558, and 2.06557 $\hbar\omega$, respectively, if
$N_{max}=16$, 18, and 20. Since the emphasis is placed at the
qualitative aspect, the convergency appears to be satisfactory.


\begin{thebibliography}{99}
\bibitem{r_EV1} V. Efimov, Phys. Lett. \textbf{B33}, 563 (1970).

\bibitem{r_EV2} V. Efimov, Sov. J. Nucl. Phys. \textbf{12}, 589 (1971).

\bibitem{r_RDE} D. E. Rutherford, Substitutional Analysis (Edinburgh
University Press, Edinburgh, 1948).

\bibitem{r_RG} G. Racah, Group Theory and Spectroscopy (Princeton University
Press, Princeton, NJ, 1951).

\bibitem{r_LEM} E. M. Lobel, Group Theory and its Applications, Vols.1,2,
and 3 (Academic Press, New York, 1968, 1971, and 1975).

\bibitem{r_EJP} J. P. Elliott, P. G. Dawber, Symmetry in Physics, Vols.1,2
(McMillan Press, London, 1979).

\bibitem{r_BCG1} C. G. Bao, Few-Body Systems \textbf{13}, 41 (1992).

\bibitem{r_BCG2} C. G. Bao and W. F. Xie, Few-Body Systems \textbf{19}, 157
(1995). 

\bibitem{r_BCG3} C. G. Bao, W. F. Xie, W. Y. Ruan, Few-Body Systems \textbf{%
22}, 135 (1997). 

\bibitem{r_MT} T. Morishita and C. D. Lin, Phys. Rev. A \textbf{64}, 052502
(2001).

\bibitem{r_PMD} M. D. Poulsen and L. B. Madsen, Phys. Rev. A \textbf{72},
042501 (2005).

\bibitem{r_EAR} A. R. Edmonds, Angular Momentum in Quantum Mechanics
(Princeton University Press, Princeton, NJ, 1957).

\bibitem{r_BTA} T. A. Brody and M. Moshinsk, Table of transformation
Brackets (Universidad Nacional Autonoma de Mexico, Monografias del Instituto
de Fisica, 1960).

\bibitem{r_BM} M. Baranger and K. T. R. Davies, Nucl. Phys. \textbf{79}, 403
(1966). 

\bibitem{r_TW} W. Tobocman, Nucl. Phys. A \textbf{357}, 293 (1981).
\end{thebibliography}
\end{document}